\documentclass[aps,prl,twocolumn,superscriptaddress,showpacs]{revtex4}

\usepackage{graphicx}% Include figure files
\usepackage{dcolumn} % Align table columns on decimal point
\usepackage{bm}      % bold math
\usepackage{epsfig}
\usepackage{amsmath,amssymb,times}
\usepackage{color}
\definecolor{orange}{rgb}{1,0.5,0}
\usepackage[normalem]{ulem}

\begin{document}

\title{Successive magnetic phase transitions and multiferroicity in Spin-1 triangular lattice antiferromagnet Ba$_3$NiNb$_2$O$_9$}

\author{J.~Hwang}
\affiliation{National High Magnetic Field Laboratory, Florida State
University, Tallahassee, Florida 32310-3706, USA}
\affiliation{Department of Physics, Florida State University,
Tallahassee, Florida 32306-3016, USA}

\author{E.~S.~Choi}
\email{echoi@magnet.fsu.edu}\affiliation{National High Magnetic Field Laboratory, Florida State
University, Tallahassee, Florida 32310-3706, USA}

\author{F.~Ye}
\affiliation{Quantum Condensed Matter Division , Oak Ridge National Laboratory, Oak Ridge, Tennessee 37381, USA}

\author{C. R. Dela Cruz}
\affiliation{Quantum Condensed Matter Division , Oak Ridge National Laboratory, Oak Ridge, Tennessee 37381, USA}

\author{Y.~Xin}
\affiliation{National High Magnetic Field Laboratory, Florida State
University, Tallahassee, Florida 32310-3706, USA}

\author{H.~D.~Zhou}
\email{hzhou10@utk.edu}\affiliation{National High Magnetic
Field Laboratory, Florida State University, Tallahassee, Florida
32310-3706, USA}\affiliation{Department of Physics and Astronomy
, University of Tennessee,
Knoxville, Tennessee 37996-1200, USA}

\author{P.~Schlottmann}
\affiliation{National High Magnetic Field Laboratory, Florida State
University, Tallahassee, Florida 32310-3706, USA}
\affiliation{Department of Physics, Florida State University,
Tallahassee, Florida 32306-3016, USA}

\date{\today}

\begin{abstract}

We report the magnetic and electric properties of Ba$_3$NiNb$_2$O$_9$, which is a quasi-two-dimensional spin-1 triangular lattice antiferromagnet (TLAF) with trigonal structure. At low $T$ and with increasing magnetic field, the system evolves from a 120 degree magnetic ordering phase (A phase) to an up-up-down ($uud$) phase (B phase) with a change of slope at 1/3 of the saturation magnetization, and then to an ``oblique'' phase (C phase). Accordingly, the ferroelectricity switches on at each phase boundary with appearance of spontaneous polarization. Therefore, Ba$_3$NiNb$_2$O$_9$ is a unique TLAF exhibiting both $uud$ phase and multiferroicity.

\end{abstract}


\maketitle

The two dimensional (2D) triangular lattice antiferromagnet (TLAF) is one of the simplest possible geometrically frustrated spin systems \cite{review2}. When the value of the spin is small, $S = 1/2$ or 1, in addition to frustration there are strong quantum spin fluctuations at low temperatures that lead to exotic ground states \cite{review1}. One  celebrated example is the unusual spin up-up-down ($uud$) ground state. Theories have predicted that quantum fluctuations should stabilize the $uud$ state in spin-1/2 TLAFs \cite{uud1,uud2,uud3}. In this collinear state, the low temperature magnetization is constant over a finite range of the magnetic field and equal to 1/3 of the saturation magnetization M$_s$. Experimentally, Cs$_2$CuBr$_4$ is a rare example of spin-1/2 TLAF, in which the $uud$ state occurs \cite{CsCuBr1,CsCuBr2}. Recently, the 1/3 M$_s$ magnetization plateau has been observed for Ba$_3$CoSb$_2$O$_9$, another TLAF with effective spin-1/2 \cite{CoSb1}. The only reported spin-1 TLAF exhibiting 1/3 M$_s$ is Ba$_3$NiSb$_2$O$_9$ \cite{NiSb}. Searching for new TLAFs with magnetization plateaus is still a very active topic in condensed matter physics.

Another interesting property of TLAFs is multiferroicity, in which the magnetic and electric degrees of freedom \cite{multi1,multi2} are strongly coupled. The multiferroicity has been observed in TLAFs with a 120 degree magnetic ordering phase for classical spins, such as ACrO$_2$ (Cr$^{3+}$ $S$ = 3/2, A = Ag and Cu) \cite{ACr1,ACr2} and RbFe(MO$_4$)$_2$ (Fe$^{3+}$ $S$ = 5/2) \cite{FeMo}. The study on RbFe(MO$_4$)$_2$ also shows that the ferroelectricity disappears in the $uud$ phase, which is stabilized by the thermal fluctuations. The theory for multiferrocity in TLAFs is still under development \cite{ACr3}, since the well established theories, such as the spin current model or the inverse Dzyaloshinshkii-Moriya model do not predict ferroelectricity in TLAFs \cite{spincurrent, Mostovoy, Dagotto}. However, it has been predicted that a broad range of trigonal materials with the 120 degree spin structure could be multiferroic, since both ACrO$_2$ and RbFe(MO$_4$)$_2$ have trigonal structure.

So far, no other TLAFs with trigonal structure and 120 degree spin structure were found to be multiferroic to further confirm this prediction. Also, no TLAFs with quantum spins ($S$ = 1/2 or 1) have been reported to show multiferroicity. Moreover, the coexistence of the $uud$ phase and multiferroicity has not been reported in any TLAF. Here, we report detailed magnetic and electric property studies on new spin-1 TLAF with  trigonal structure, Ba$_3$NiNb$_2$O$_9$ (Ni$^{2+}$ has $S$ = 1). We observed successive magnetic phase transitions at 1/3 and $\sqrt{3}/3$ of the saturation magnetization, which are also accompanied by spontaneous reversible polarization and dielectric anomalies. Although a distinctive magnetization plateau was not observed due to the polycrystallinity and the finite temperature, our results strongly suggest that the title compound is a rare example of TLAF with coexistence of collinear magnetic structure and multiferroicity.

\begin{figure}[tbp]
\linespread{1}
\par
\begin{center}\includegraphics[trim=0.5in 2.8in 0.5in 0.5in, clip=true,width=3.2in]{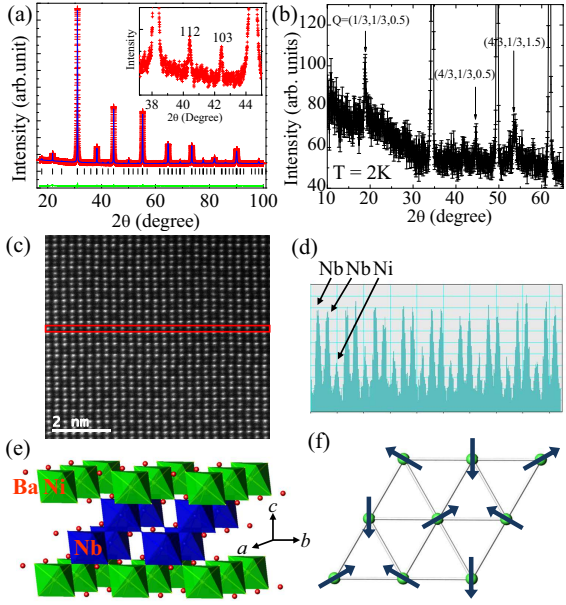}
\end{center}
\par
\caption{(Color online) (a) The room temperature XRD pattern (Cu $K_{\alpha1}$ $\lambda$ = 1.54059 {\AA}) for Ba$_3$NiNb$_2$O$_9$ (plus marks). The solid curve is the best fit from the Rietveld refinement using FullProf. Inset: the (112) and (103) peaks from the XRD pattern. (b) The neutron diffraction pattern measured at 2 K showing several magnetic Bragg reflections. (c) Atomic resolution STEM HAADF Z-contrast image looking down [010]. Ba, Nb, and Ni atoms show brightest, brighter, and weak contrast, respectively. (d) Line profile of the atomic columns intensity indicated by the box in (c) showing the ordering sequence of -Ni-Nb-Nb-.(e) Schematic crystal structures for Ba$_3$NiNb$_2$O$_9$: the green octahedra represent Ni sites and the blue octahedra represent Nb sites. (f) Zero field magnetic structure below $T_{\text{N1}}$ composed of Ni$^{2+}$ ions and the arrows representing their spins.}
\end{figure}

Polycrystalline Ba$_3$NiNb$_2$O$_9$ samples were prepared by solid state reaction. Appropriate mixtures of BaCO$_3$, NiO and Nb$_2$O$_5$ were ground together, pressed into pellets, and then calcined in air at 1230 $^{\circ}$C for 24 hours. The room temperature powder x-ray diffraction (XRD) pattern [Fig.~1(a)] of the as-prepared sample can be indexed as a hexagonal P-3m1 structure with $a$ = $b$ = 5.7550(5) {\AA} and $c$ = 7.0656(2) {\AA} as reported previously \cite{NiNb1}, which belongs to the trigonal space group. The 1:2 ordering of the B site ions (-Ni-Nb-Nb-) in this structure has been confirmed by two evidences: (i) the appearance of low intensity Bragg peaks in the XRD pattern related to the ordering structure \cite{NiNb2}, such as the (112) and (103) peaks [inset of Fig.~1(a)] and (ii) the high angle annular dark field (HAADF) Z-contrast image from scanning transmission electron microscopy (STEM) [Figs.~1(c and d)]. The intensity of the Nb and Ni atom columns in the image clearly shows the -Ni-Nb-Nb- ordering viewed down the [010] direction due to their atomic number Z difference \cite{Pennycook, Loane}. No defects are seen on the Ni or Nb sites. We noticed that with the annealing temperature above 1230 $^{\circ}$C, the sample will turn to be Ba(Ni$_{0.33}$Nb$_{0.67}$)O$_3$ with B-site disorder and Pm-3m cubic structure. With the P-3m1 structure, Ba$_3$NiNb$_2$O$_9$ can be represented as a framework consisting of corner-sharing NiO$_6$ and NbO$_6$ octahedra [Fig.~1(e)]. The Ni ions occupy the 1b Wyckoff sites, and this site forms the triangular lattice in the $ab$ plane [Fig.~1(f)]. Therefore, the structure can be regarded as a pseudo-two-dimensional triangular magnet, i.e., the Ni magnetic triangular lattices are magnetically separated by the two non-magnetic Nb layers.

The susceptibility $\chi$ measured at $H$ = 0.1 T [Fig.~2(a)] shows a peak around 4.9 K related to the long range magnetic ordering. The linear Curie-Weiss fit of 1/$\chi$ at high temperature (not shown here) gives a $\theta_{\text{CW}}$ = -16.4 K, showing the antiferromagnetic nature of the exchange interactions and an effective moment $\mu_{\text{eff}}$ = 3.15 $\mu_{\text{B}}$/Ni, which corresponds to a Land\'e $g$-factor $g$ = 2.23 and is typical for Ni$^{2+}$ ions \cite{Ni}. Assuming that each Ni$^{2+}$ spin only interacts via exchange $J$ with the $z=6$ nearest neighbor spins on the triangular lattice, the Heisenberg Hamiltonian is $J \sum_{<i,j>}$S$_{i}\cdot$S$_{j}$. In mean-field theory, this leads to a Curie-Weiss law with $\theta_{\text{CW}}=-zJ$S(S+1))/3k$_{\text{B}}$. For Ba$_3$NiNb$_2$O$_9$, $S$ = 1 and $J$/k$_{\text{B}}$ = -1/(4$\theta_{\text{CW}}$) = 4.1 K. The neutron diffraction pattern measured at 2 K [Fig.~1(b)] shows magnetic Bragg peaks at Q = ($n_1$ + 1/3, 1/3, $n_2$ + 1/2) ($n_i$: integer). Accordingly, its refinement (not shown here) reveals a magnetic structure with collinear AFM spins between the nearest neighbor layers and the 120 degree AFM ordering within the triangular lattice with an ordered moment of 1.8 $\mu_{\text{B}}$ for each Ni$^{2+}$ ion. This refinement also shows a B sites ordered P-3m1 lattice structure at 2 K, which means there is no obvious structural distortion below 4.9 K.

For $H >$ 5 T, the peak of the susceptibility $\chi$ becomes broader [Fig. 2(a)] and the related d$\chi$/d$T$ vs. $T$ curves display two peaks [inset of Fig. 2(a)]. We define $T_{\text{N1}}$ and $T_{\text{N2}}$ by the locations of the higher and lower temperature peaks, respectively. Below 9 T, both $T_{\text{N1}}$ and $T_{\text{N2}}$, decrease with increasing field. The zero field specific heat data [Fig.~2(b)] shows a single peak around $T_{\text{N1}}$. For $H >$ 5 T, this peak becomes broader and could be regarded as two peaks as indicated by lines in Fig.~2(b) for 9 T. Their locations are consistent with those of $T_{\text{N1}}$ and $T_{\text{N2}}$ defined from d$\chi$/d$T$.

\begin{figure}[tbp]
\linespread{1}
\par
\begin{center}
\includegraphics[width=3.4in]{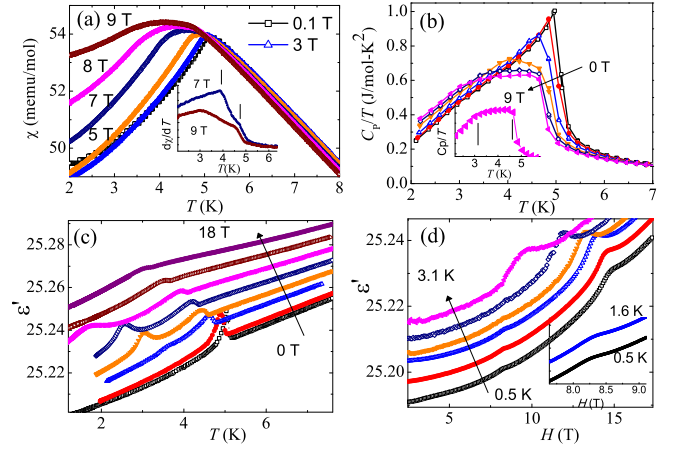}
\end{center}
\par
\caption{(Color online) (a) Temperature dependence of $\chi$ for Ba$_3$NiNb$_2$O$_9$ at different fields. Inset: d$\chi$/d$T$ $vs.$ $T$ for $H$ = 7 and 9 T. (b) Temperature dependence of $C$$_{\text{P}}$/$T$ at 0, 3, 5, 7, 8, and 9 T. Inset: Expansion of the 9 T data. (c) Temperature dependence of $\varepsilon^\prime$ at 0, 4, 8, 10, 12, 14, 16 and 18 T, and (d) magnetic field dependence of $\varepsilon^\prime$ at 0.5, 0.9, 1.6, 2, 2.6 and 3.1 K. Inset: Expansion around $H$ = 8 T for 0.5 and 1.6 K (data is offset for clarity). The vertical lines in (a) and (b) indicate $T_{\text{N1}}$ (higher $T$) and $T_{\text{N2}}$ (lower $T$).}

\end{figure}

Samples with rectangular shape with typical dimensions of 5.0$\times$4.0$\times$0.15 mm$^3$ were cut from the pressed pellet to measure the dielectric constant and the pyroelectric current using the techniques found elsewhere \cite{Hwang}. The real part of the dielectric constant ($\varepsilon^\prime$) measured at zero field [Fig.~2(c)] shows a sharp peak at $T_{\text{N1}}$. With increasing field, this peak shifts to lower temperatures and becomes broader. For $H >$ 8 T, a second peak well separated from the first peak appears at lower temperatures. The field dependence of $\varepsilon^\prime$ measured at 0.5 K shows two peaks around 8 T and 15 T, respectively [Fig.2(d)]. With increasing temperature, the peak around 8 T does not move significantly while the peak around 15 T moves to lower fields. The two peaks tend to merge at about 3 K.

\begin{figure}[tbp]
\linespread{1}
\par
\begin{center}
\includegraphics[width=3.4in]{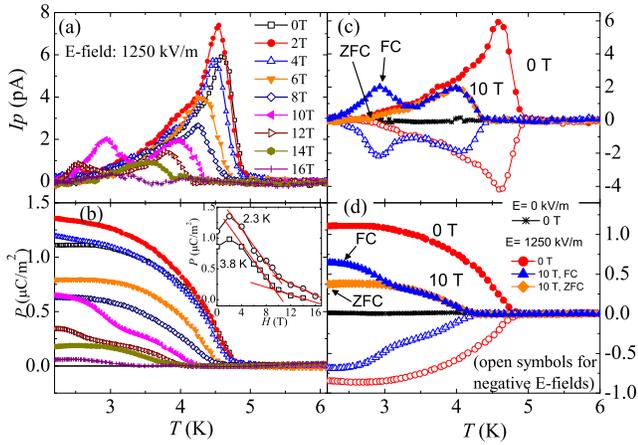}
\end{center}
\par
\caption{(Color online) Pyroelectric current ($I_p$) (a) and polarization (b) as a
function of temperature at different magnetic fields under positive electric poling field
and FC conditions. The inset of (b) shows polarization vs. magnetic field at 2.3 K and 3.8 K under
ZFC condition. $I_p$ (c) and polarization (d) at $H$ = 0 and 10 T under positive (solid symbols) and
negative (open symbols) poling fields and ZFC and FC conditions.}

\end{figure}

The pyroelectric current was measured on warming after poling the crystal in an electric field while cooling down from above $T_{\text{N1}}$ with magnetic field (FC) or without magnetic field (ZFC). The spontaneous polarization was obtained by integrating the pyroelectric current with respect to time. At zero magnetic field, the pyroelectric current, and accordingly the spontaneous electric polarization, begin to develop below $T_{\text{N1}}$ [Fig.~3(a)]. With opposite poling electric field, the pyroelectric current direction can be reversed [Fig.~3(c)].  This indicates the ferroelectric (FE) nature of the ground state, which is clearly coupled with the 120 degree magnetic ordering. The obtained polarization is around 1.1 $\mu$C/m$^2$ at 0 T. The polarization overall decreases and shifts to lower temperatures with increasing magnetic field [Fig.~3(b)], which is consistent with the behavior observed for the higher temperature peak in $\varepsilon^\prime$. For the polarization emerging at $T_{\text{N1}}$, there is no difference between the ZFC and FC measurements.

For $H \geq$ 8 T another peak of the pyroelectric current (or another step-like increase of the polarization) is found at the same temperature where the lower temperature peak of $\varepsilon^\prime$ is observed. Note that this low temperature polarization increase is observed only when the sample is cooled under the presence of {\it both}, magnetic and electric, fields, i.e., after magnetoelectric (ME) annealing. For example, the lower temperature peak for the pyroelectric current disappears for the ZFC measurement, as seen in Fig.~3(c) for the 10 T data. For some ME systems, the reversible pyroelectric peak is also observed under ME annealing without invoking a ferroelectric transition as a consequence of the combined effects of magnetic domains and the ME effect \cite{Rado2, Shtrikman,Hwang}. Therefore, the additional increase of the polarization and the low temperature $\varepsilon^\prime$ anomaly are more likely to arise from a magnetoelectric mechanism, rather than from another FE transition with a different order parameter.

\begin{figure}[tbp]
\linespread{1}
\par
\begin{center}
\includegraphics[width=3.4in]{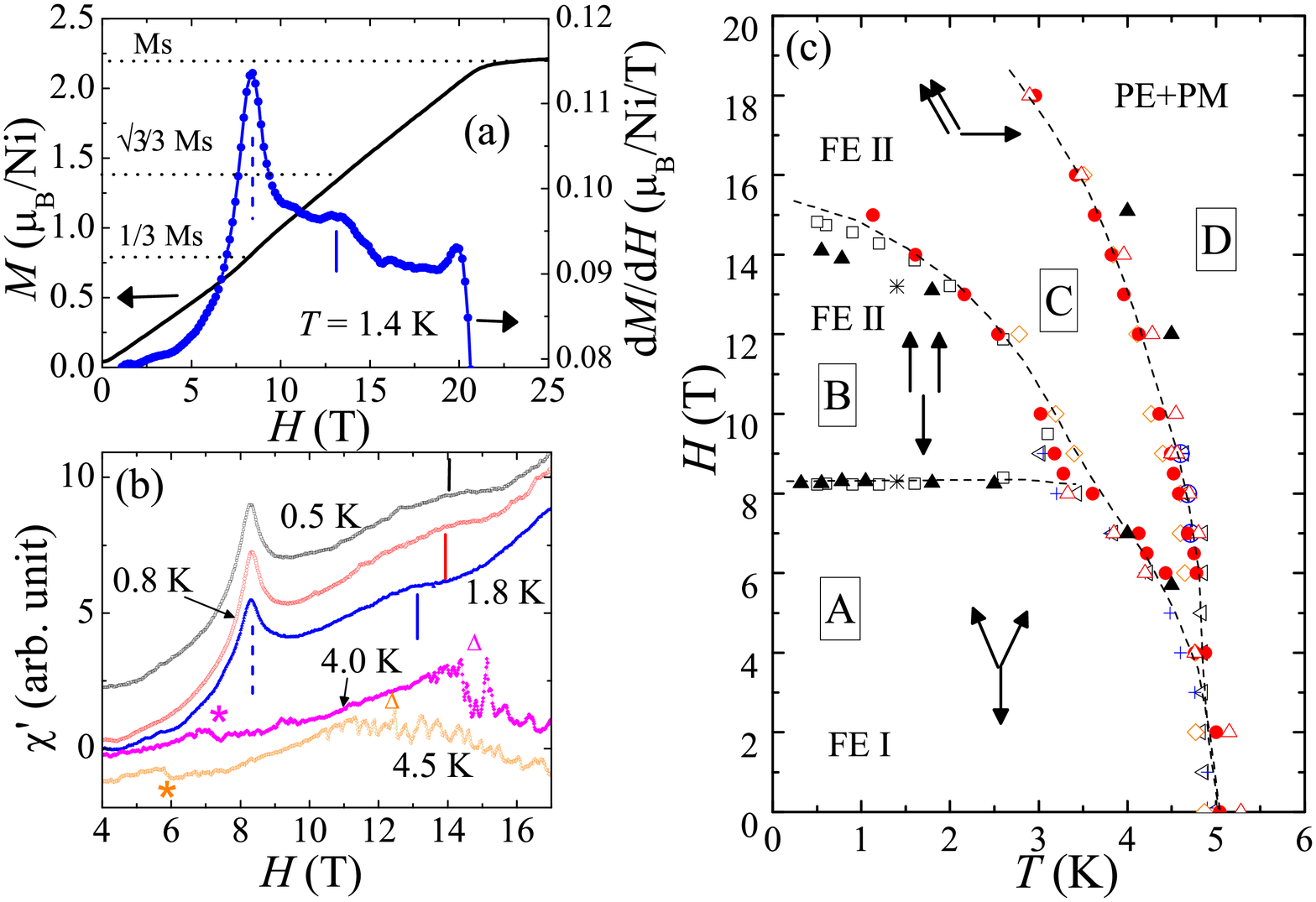}
\end{center}
\par
\caption{(Color online) (a) DC magnetization ($M$) and its derivative (d$M$/d$H$) for $T =$ 1.4 K. (b) AC susceptibility as a function of magnetic field $H$ for four temperatures. The asterisks and the triangles indicate A-C and C-D phase boundary respectively. (c) $H$-$T$ phase diagram for Ba$_3$NiNb$_2$O$_9$. Different symbols denote phase boundaries obtained from different techniques. $\Box$: $\varepsilon^\prime$ $H$-sweep, {\color{red} $\bullet$}: $\varepsilon^\prime$ $T$-sweep, {\color{red} $\diamond$}: polarization, $\lhd$: $C$$_{\text{P}}$, $\blacktriangle$: AC susceptibility $H$-sweep, {\color{red} $\triangle$}: AC susceptibility $T$-sweep, {\color{blue}$\oplus$}: DC susceptibility $T_{\text{N1}}$, {\color{blue}+}: DC susceptibility $T_{\text{N2}}$, $\ast$: DC magnetization. The arrows indicate spin structures.The dashed (solid) vertical lines in (a) and (b) indicate the
transition fields to $uud$ (oblique) phases. The dashed lines in (c) are guides to the eyes.}
\end{figure}

Combining all the values for the transition temperatures and critical magnetic fields observed from the various techniques described above, the low temperature $H$-$T$ phase diagram for Ba$_3$NiNb$_2$O$_9$ shown in Fig.~4(c) is obtained. Below 4.9 K and with increasing magnetic field, at least three magnetic phases (labeled as A, B, and C in Fig.~4(c)) are consistently observed. In the A phase, the spins form the 120 degree coplanar AFM ordering at zero field. Under small external magnetic fields, the order in the A phase is expected to gradually deviate from the 120 deg structure while maintaining a similar degree of continuous degeneracy \cite{Kawamura}. At higher fields, as the theories predict \cite{uud1,uud2,uud3}, for a AFM triangular lattice, the quantum fluctuations for small spins can lift the classical degeneracy in the A phase to select the B phase. We tentatively assign the $uud$ state with a 1/3 M$_s$ magnetization plateau stabilized in a finite magnetic field range to the B phase \cite{uud1,uud2,uud3}.

In order to examine the possibility of the magnetization plateau in the B phase, the DC magnetization and AC susceptibility ($\chi^\prime$) on were measured. As seen in Fig.~4(a), the DC magnetization saturates around 23 T with $M_s$ = 2.2 $\mu_{\text{B}}$. A slope change of $M$ vs.~$H$ around 8 T is observed, at which the magnetization is 0.77 $\mu_{\text{B}}$ $\approx$ 1/3 $M_s$. Experimentally, the d$M$/d$H$ curve shows a distinctive feature characterized by a broad valley in the magnetization plateau region and a sharp peak and a broad shoulder respectively on entering and exiting the putative plateau region \cite{CoSb1,NiSb,Svistov}. In Ba$_3$NiNb$_2$O$_9$, the most distinctive feature (valley) is not apparent in the d$M$/d$H$ curve as a consequence of the lack of the plateau-like magnetization. On the other hand, the $\chi^\prime$ measurement, being a technique to probe d$M$/d$H$ directly, revealed all the features as shown in Fig.~4(b): a dominant peak at 8 T followed by a broad valley at 9 T and a broad shoulder like peak at around 14 T. However, the $\chi^\prime$ at the valley is not the smallest as the overall $\chi^\prime$ increases with field. There are two possible reasons why the magnetization plateau was not clearly observed: (a) the polycrystalline nature of our sample and (b) that even in a single crystal the plateau is manifested only in certain directions of the field with respect to the crystalline axes.

The 14 T feature is also seen in d$M$/d$H$ as a broad peak, where the magnetization is approximately $\sqrt{3}/3$ $M_s$ [Fig.~4(a)]. This suggests the 14 T feature is due to the transition from the $uud$ to a coplanar 2:1 canted "oblique" phase (C phase) with one spins in the A phase rotated to be parallel with another spin, which gives $\sqrt{3}/3$ $M_s$. We emphasize here that our spin structure model in the oblique phase is based on the magnetization data and more rigorous experiments like neutron diffraction should be performed to confirm the structure. In general, the spin structure in the oblique phase, a generic term for a spin structure with two parallel spins tilted with respect to the third, is unique to each TLAF system depending on the magnitude of the spin and hence on the order of the fluctuations. It is noteworthy that our suggested 2:1 canted spin structure is the same as that of RbFe(MO$_4$)$_2$ under 10 T in-plane magnetic field, which was confirmed by neutron experiments (see Fig. 1(d) of Ref. \cite{FeMo}). At higher temperatures, the $\chi^\prime$ data also show features (indicated as asterisks and triangles in Fig.4(b)) at the phase A-C and C-D (paramagnetic phase) phase boundaries. These experimental observations clearly suggest that Ba$_3$NiNb$_2$O$_9$ evolves through successive magnetic phase transitions from the A phase to $uud$ (B) phase and then to oblique (C) phase with increasing magnetic field. Whether a magnetization plateau should exist in the B phase is still an open question. Experiments performed on single crystal samples with in-plane magnetic fields at lower temperatures are expected to reveal the existence of the magnetization plateau more clearly.

For TLAFs with classical spins, thermal fluctuations stabilize the $uud$ phase at finite temperatures showing a plateau-like magnetization, hence the magnetic field range of the $uud$ phase decreases with decreasing temperature. In Ba$_3$NiNb$_2$O$_9$, the $uud$ phase has a range of about 8 T even at 0.5 K ($\ll$ $T_{\text{N1}}$) and the width of the $uud$ phase increases with decreasing temperature. This suggests that the $uud$ phase has a quantum mechanical origin due to the small spin $S$ = 1 for Ni$^{2+}$. This makes Ba$_3$NiNb$_2$O$_9$ a rare $S$ = 1 TLAF showing a $uud$ phase driven by quantum fluctuations. Another noteworthy feature in the phase diagram of Ba$_3$NiNb$_2$O$_9$ is that within a narrow temperature window, 3.5 K $<$ $T$ $<$ 4.5 K, with increasing field the spin configuration goes from the A phase to the C phase directly without passing through the B phase (see online supplemental material for details). This is significantly different from the other TLAF systems with $uud$ phase mentioned in the introduction, either for classical or quantum spins. For those systems, in order to reach the C phase from the A phase with increasing field, it is always necessary to pass through the B phase. To our knowledge, so far no theory predicts this feature and we leave the explanation for it as an open question for future investigations.

More interestingly, as the polarization measurements show, all these three magnetic phases have a FE ground state. Studies of multiferroic properties of TLAFs have so far been limited to systems with classical spins. Two celebrated examples are (i) ACrO$_2$ (A = Cu and Ag) with trigonal {\it R-3m} structure and (ii) RbFe(MO$_4$)$_2$ with trigonal {\it P-3m} structure. It has been proposed that the ferroelectricity in ACrO$_2$ is driven by the helical spin-spiral order of the Cr$^{3+}$ ($S$ = 3/2) spins among the stacked Cr-layers \cite{ACr3}. The zero field magnetic structure of RbFe(MO$_4$)$_2$ has AFM spins between the nearest neighbor layers and a 120 degree AFM ordering within the triangular layers. A phenomenological model based on symmetry arguments (the broken inversion symmetry and the absence of a mirror plane perpendicular to the $c$ axis) was proposed to explain the ferroelectricity in the A phase for RbFe(MO$_4$)$_2$ \cite{FeMo}. A similar situation could lead to the ferroelectricity in the A phase for Ba$_3$NiNb$_2$O$_9$, since the two Nb layers between the adjacent Ni layers lead to the absence of a mirror plane perpendicular to the $c$ axis [Fig. 1(e)]. Although a smaller polarization, P = 1.1 $\mu$C/m$^2$, was observed due to the polycrystalline nature of Ba$_3$NiNb$_2$O$_9$, its value is still comparable to that of RbFe(MO$_4$)$_2$ obtained with single crystal samples.

A difference between RbFe(MO$_4$)$_2$ and Ba$_3$NiNb$_2$O$_9$ is the electric ground state at higher magnetic fields. In RbFe(MO$_4$)$_2$, the ferroelectricity disappears in the B phase ($uud$ phase), which was interpreted as the inversion symmetry in the triangular $uud$ magnetic structure being reinstated. In contrast, the ferroelectricity persists in the $uud$ and the oblique phases of Ba$_3$NiNb$_2$O$_9$. The inset of Fig. 3(b) shows the magnetic field dependence of the polarization at 2.3 K and 3.8 K obtained from the pyroelectric current measurement under ZFC. The polarization increases initially up to $H$ = 2 T followed by a monotonous decrease with the increasing magnetic field up to 9 T. Then, there is a change of curvature around 9 T, above which the polarization is larger than the projected value extrapolated from the linear behavior of the lower magnetic fields. This behavior suggests that the mechanism for the FE state in B and C phases is different from that for the A phase under lower magnetic fields.

One possible explanation for the FE phase at high magnetic fields in Ba$_3$NiNb$_2$O$_9$ is that the inversion symmetry is broken by deforming the triangular lattice via magnetostriction. In the triangular spin structure, magnetostriction driven by spin correlation ({\bf S}$_i$ $\cdot$ {\bf S}$_j$) cannot break the inversion symmetry, but the anisotropic magnetostriction driven by external magnetic fields can. In the distorted triangular lattice, the space inversion and the three fold rotation are not symmetric operations any more, which can lead to finite polarization in the $ab$ plane (see online supplemental material for details). Magnetostriction often plays an important role in inducing magnetic multiferroicity \cite{DyFeO3,Choi}. The polarization values close to the B-C phase boundary at fixed $T$ are too small to draw any conclusions, but the continuous polarization increase with decreasing temperature while crossing the B-C boundary at fixed magnetic fields under the ZFC condition [Fig.~3(d)] suggests that the FE order parameter is same in the B and C phases. From these observations, we speculate that the magnetostriction starts to play an important role in inducing ferroelectricity at high magnetic fields (B and C phases), while the spin chirality in the A phase is the driving force of the multiferroicity at low magnetic fields.

In summary, although more work is needed to confirm the magnetic structure of the A, B and C phases, the detailed magnetic and electric studies on Ba$_3$NiNb$_2$O$_9$ reveal that the system (i) is a new spin-1 TLAF system showing two magnetic phase transitions bracketing an intermediate up-up-down phase driven by quantum fluctuations; (ii) is a multiferroic system with FE ground states stabilized at all three magnetic phases. These findings show that Ba$_3$NiNb$_2$O$_9$ is a rare TLAF with strong couplings between the successive magnetic phase transitions and the ferroelectricity. One step further, we also confirmed the prediction that a broad range of trigonal materials with the 120 degree spin structure can be multiferroic.

\begin{acknowledgments}
This work was supported by NSF-DMR-0654118 and the State of Florida. P.~S.~is supported by the DOE under Grant No.~DE-FG02-98ER45707.
\end{acknowledgments}

\end{document}